\documentclass{pazhb}
\usepackage[hyperfootnotes=false]{hyperref}
\usepackage{color}
\usepackage{epsf}
\usepackage[utf8]{inputenc}
\usepackage[russian]{babel}
\usepackage[T2A]{fontenc}
\usepackage{amsmath}
\usepackage{amsfonts}
\usepackage{amssymb}
\usepackage{epsf}
\usepackage{graphicx}
\usepackage{gensymb}
\usepackage{float}
\usepackage{longtable}
\usepackage{changes}
\usepackage{wrapfig}
\usepackage{comment}
\newcommand{\obj}{Gaia23cer}
\DeclareMathOperator*{\argmin}{argmin}

\begin{document}

\journalinfo{2024}{50}{5}{335}[349]
\UDK{///}

\title{Spectroscopic and photometric study of the new eclipsing polar Gaia23cer}
\author{%\bf \hspace{-1.3cm}\copyright\, 2020 г. \ \
A.I. Kolbin\address{1,2,5}\email{kolbinalexander@mail.ru}, 
E.P.~Pavlenko\address{3}\email{kolbinalexander@mail.ru (A.I. Kolbin), eppavlenko@gmail.com (E.P.~Pavlenko)},
V.Yu.~Kochkina\address{1,2}, 
A.S.~Vinokurov\address{1}, 
S.Yu.~Shugarov\address{4,6}, 
A.A.~Sosnovskij\address{3}, 
K.A.~Antonyuk\address{3}, 
O.I.~Antonyuk\address{3}, 
N.V.~Pit\address{3}, 
M.V.~Suslikov\address{1,2}, 
E.K.~Galimova\address{2}, 
N.V.~Borisov\address{1}, 
A.N.~Burenkov\address{1}, 
O.I.~Spiridonova\address{1}
\addresstext{1}{Special Astrophysical Observatory of RAS, 357147, Nizhnii Arkhyz, Karachai-Cherkessian Republic, Russia}
\addresstext{2}{Kazan (Volga Region) Federal University, 420000, Kazan, Russia}
\addresstext{3}{Crimean Astrophysical Observatory, 298409, Nauchnyi, Crimea, Russia}
\addresstext{4}{Astronomical Institute, Slovak Academy of Sciences, Tatranska Lomnica, Slovakia}
\addresstext{5}{North-Caucasus Federal University, 355017, Stavropol, Russia}
\addresstext{6}{Sternberg Astronomical Institute, Moscow State University, 119234, Moscow, Russia}
}

%\vspace{2mm}
%\received{\today}
%\sloppypar
%\vspace{2mm}
%\noindent

\shortauthor{Kolbin, Pavlenko, et al.}

\shorttitle{The study of the new eclipsing polar Gaia23cer}

\begin{abstract}

We present the results of the optical study of the new eclipsing polar Gaia23cer. We analyzed the brightness variability of the polar in high ($\langle r \rangle \approx 16.5\mathrm{\,mag}$) and low  ($\langle r \rangle \approx 19.2\mathrm{\,mag}$) states. The system has an orbital period $P_{orb} = 102.0665 \pm 0.0015$~min and exhibits deep eclipses with a duration $\Delta t_{ecl} = 401.30 \pm 0.81$~s. The spectra have a red cyclotron continuum with the Zeeman H$\alpha$ absorption triplet forming in a magnetic field with a strength of $15.2 \pm 1.1$~MG. The source of emission lines has a high radial velocity semi-amplitude ($K\approx 450$~km/s) and its eclipse lags behind the eclipse of the white dwarf. The mass $M_1=0.79 \pm 0.03 M_{\odot}$ and temperature $T=11350 \pm 650 K$ of the white dwarf have been found by modelling the spectral energy distribution.  The eclipse duration corresponds to a donor mass $M_2 = 0.10-0.13M_{\odot}$ and an orbital inclination $i=84.3-87.0^{\circ}$. The donor temperature was estimated to be $T\approx 2900K$ by modelling the elliptical variability and eclipse depth.

\end{abstract}

%***************************************************************
\section{Introduction}

Polars (or AM Her stars) are close binary systems that consist of a strongly magnetized ($B\sim 10-100$ MG) white dwarf and a low-mass cool star (usually an M-dwarf) filling its Roche lobe. The strong magnetic field of polars prevents the formation of an accretion disk by channeling the accreted material along magnetic field lines to the vicinity of the magnetic poles. When the infalling gas interacts with the surface of the white dwarf, hot ($T\sim10$~keV) accretion spots are formed. They are sources of X-ray bremsstrahlung and polarized cyclotron radiation in the optical and infrared ranges. The strong magnetic field also makes the polars synchronous systems, where the rotation period of the white dwarf is equal to the orbital period. For more details on AM Her stars, see \cite{Cropper90}.

The studies of AM Her stars are important in several aspects. First, the study of polars is necessary to understand the origin and structure of the magnetic fields of white dwarfs in binary systems \citep{Belloni20, Schreiber21}. Second, because of the large magnetospheric radius comparable to the semimajor axis, AM~Her systems are convenient for studying the physics of the interaction of the accreted gas with the magnetic fields of the accretors \citep{Hameury86, Li99}. Third, the study of polars is important for understanding the evolution of close binary systems with magnetized components \citep{Belloni20}. For this purpose, the study of eclipsing systems is especially important. The presence of eclipses makes it possible to determine reliable parameters of a binary system (the component masses, the orbital inclination). In addition, by modeling the eclipse profile at the polars, the geometry of the accretion flow can be reconstructed (see, e.g., \cite{Harrop99, Harrop01}).

The object Gaia23cer (ZTF18abunixr, AT2023row;  $\alpha = 01^h 26^s 07.79^m$, $\delta = +12^{\circ} 10' 48.94''$ (J2000)) was detected as an optical transient from Gaia observations \cite{Hodgkin23}. \cite{Simon23} detected its photometric variability with a period of 0.065(6)~day and deep eclipses. \cite{Sosnovskij23} performed multiband photometric observations of Gaia23cer. Double-humped light curves with deep eclipses ($\Delta B \approx \Delta V \approx \Delta R_C \approx 5\mathrm{\,mag}$ и $\Delta I_C \approx 3\mathrm{\,mag}$) of duration $\approx 7.6$~min are reported. Based on the analysis of Gaia23cer photometry, the authors suggested that the source can be classified as an AM~Her-type star.

In this paper we perform an optical study of Gaia23cer using our photometric and spectroscopic observations. In Section 2 we describe our observations and the data reduction methods. Next, in Section 3, we analyze the variability of the polar on timescales of years and hours. In Section 4 we analyze the spectroscopic behavior of Gaia23cer. The parameters of the white dwarf and the donor are determined in Sections 5 and 6, respectively. In the Conclusions we summarize our results.

\section{2. Observations and data reduction}

\subsection{Spectroscopy}

A set of spectra for Gaia23cer was obtained with the 6-m BTA telescope at the Special Astrophysical Observatory of the Russian Academy of Sciences (SAO RAS) on the nights of September 12/13 and October 19/20, 2023. The observations were carried out with the SCORPIO-1 focal reducer\footnote{A description of the SCORPIO-1 instrument can be found at https://www.sao.ru/hq/lsfvo/devices/scorpio/scorpio.html.} in the mode of long-slit spectroscopy \citep{Afan05}. A log of spectroscopic observations is given in Table \ref{log_spec}.

On the first night we used a volume phase holographic grating, VPHG1200B (1200 lines/mm), with the $3700-5300$~\AA\, range coverage. We obtained ten spectra near the eclipse phase\footnote{The fourth exposure is not used in the subsequent analysis, since it coincided with the deepest part of the eclipse and the object is not detected in the spectrum.} with an exposure of 120~s at and slit width of $1.2''$. The observations were performed in good astroclimatic conditions with a seeing of $\approx 2.3''$.

The October 19/20, 2023 observations were carried out with VPHG1200G grism (1200 lines/mm) and a $1.2''$-wide slit covering the $3900 - 5700$~\AA\, range with a resolution $\Delta \lambda \approx 5$~\AA. On the same night we took four spectra with VPHG550G grism (550 lines/mm) with the range coverage $4000-7200$~\AA\, and a resolution $\Delta \lambda \approx 8$~\AA\, (the slit width was also  $1.2''$). The observations were carried out in poor weather conditions with slight cloudiness and high ($\mathrm{FWHM} \approx 5''$) seeing. The spectrograph slit was oriented along the line connecting Gaia23cer and the neighboring bright star (Gaia DR2 2586433433115244288, $\mathrm{G}=16.76\mathrm{\,mag}$). Since the neighboring star is close to Gaia23cer (a distance of $\approx 4.2''$), the spatial profiles of these sources overlapped in conditions of a highly turbulent atmosphere.

The spectroscopic data were reduced using the IRAF\footnote{The IRAF astronomical data processing and analysis package is accessible at https://iraf-community.github.io.} and MIDAS software packages. The spectra images were bias subtracted, flat-fielded and were wavelength calibrated using Th-Ar lamp frames. The cosmic rays were removed using the LACosmic code \citep{Dokkum}. Because of the poor seeing on October 19/20, the profile of Gaia23cer overlapped with the profile of the neighboring bright star. To separate the spectra of the two sources, their profiles along the slit were fitted by the sum of two Gaussians. The separation of the Gaussians was fixed at a value corresponding to the angular distance between the sources. The widths of the Gaussians were assumed to be identical and to change gradually with wavelength. The fluxes from the star were measured from the area of the fitted Gaussians. The observations were carried out under variable cloudiness and the extracted fluxes may contain significant errors. For this reason, we analyzed only the shape of the spectra. We also calculated the barycentric Julian dates (BJDs) and the barycentric radial velocity corrections.

\subsection{Photometry}

The photometric observations of Gaia23cer were carried out from September 2023 to January 2024 with several telescopes: the 2.6-m ZTSh telescope and 1.25-m AZT-11 telescope at the Crimean Astrophysical Observatory of the Russian Academy of Sciences, the 60-cm Zeiss-600 telescope (G2; Tatranska Lomnica, Slovakia), and the 1-m Zeiss-1000 telescope at SAO RAS. They were equipped with CCD photometers with filters of the Johnson–Cousins photometric system. Additional observations were performed without using any filters with the K-380 telescope at the Crimean Astrophysical Observatory. A log of photometric observations of Gaia23cer is presented in Table \ref{log_phot}.

The CCD frame processing included bias subtraction, flat fielding, and cosmic-ray removal. For the $\mathrm{I_C}$-band images we also eliminated the fringes. Because of the presence of a bright neighboring star, the fluxes from Gaia23cer were measured by the method of PSF photometry implemented in the photutils library\footnote{The photutils library is accessible at https://photutils.readthedocs.io/en/stable/.}.

\begin{table*}
\caption{
Log of spectroscopic observations of Gaia23cer with BTA/SCORPIO. The dates of observations, the number of spectra obtained, the grisms used, the spectral ranges, the spectral resolutions, and the exposures ($\Delta t_{exp}$) are listed.
}
\label{log_spec}
\begin{center}
\begin{tabular}{lccccc}
\hline
 Date (UT)    & $N$	  & Grism & Диапазон, \AA & $\Delta \lambda$, \AA & $\Delta t_{exp}$, sec \\
\hline
12/13 сен. 2023 г. & 10 & VPHG1200B & 3700 -- 5300 & 5.5 & 120  \\
19/20 окт. 2023 г. & 17 & VPHG1200G & 3900 -- 5700  & 5 & 300  \\
19/20 окт. 2023 г. & 4 & VPHG550G & 4000 -- 7200 & 8 & 300  \\
\hline
\end{tabular}
\end{center}
\end{table*}

\begin{table*}
\caption{
 Log of photometric observations of Gaia23cer. The telescopes and CCD detectors involved in the observations, the duration of our observations, the number of images obtained ($N$), the photometric bands (``integral'' is the observations without photometric filters), and the exposures ($\Delta t_{exp}$) are listed.}
\label{log_phot}
\begin{center}
\begin{tabular}{lccc||lccc}
\hline
 Time interval,    & $N$	  & Filter & $\Delta t_{exp}$ &  Time interval,    & $N$	  & Filter & $\Delta t_{exp}$, \\
 BJD-2459000          &     &        & sec & BJD-2459000          &     &        & sec \\\hline
 \multicolumn{4}{c||}{ЗТШ/ELSE-i 1K$\times$1K BI MID} & 1215.540  & 1  & $\mathrm{U}$&825\\
  1201.427--1201.589& 83 & $\mathrm{B}$   &30 & 1215.507--1215.566& 6  & $\mathrm{V}$ &60 \\
  1201.427--1201.590& 81 & $\mathrm{V}$   &30 & 1215.500--1215.619& 43 & $\mathrm{R_C}$&150 \\
  1201.427--1201.591& 84 & $\mathrm{R_C}$ &30 & 1215.510--1215.569& 13 & $\mathrm{I_C}$&30 \\
  1201.426--1201.590& 85 & $\mathrm{I_C}$ &30 & 1235.349--1235.576& 53& $\mathrm{B}$&180\\
  1229.287--1229.399& 283& $\mathrm{I_C}$ &30 & 1235.317--1235.346& 17& $\mathrm{V}$&150\\
  1230.349--1230.430& 99 & $\mathrm{B}$   &60 & 1236.304--1236.522& 43 & $\mathrm{B}$&180 \\
  1258.356--1258.532& 239& $\mathrm{I_C}$ &60 & 1236.302--1236.308& 2 & $\mathrm{V}$&120 \\
 \multicolumn{4}{c||}{Цейсс-1000/EEV 42-40 ($2\mathrm{K}\times2\mathrm{K}$)}         & 1236.300--1236.419& 72& $\mathrm{R_C}$&120 \\
  1201.487--1201.585& 153& $\mathrm{V}$   &20 & 1339.198--1339.302& 35& $\mathrm{V}$&300\\
  1202.415--1202.485& 120& $\mathrm{I_C}$ &20 & \multicolumn{4}{c}{АЗТ-11/ProLine PL23042} \\
  1231.315--1231.423& 168& $\mathrm{I_C}$ &20 & 1207.267--1207.373& 49& $\mathrm{I_C}$&180\\
  \multicolumn{4}{c||}{Цейсс-600/FLI ML 3041}  & 1208.256--1208.393& 64& $\mathrm{I_C}$&180\\
  1203.412--1203.558& 19 & $\mathrm{B}$   &60--150 & 1209.248--1209.411& 75& $\mathrm{I_C}$&180\\
  1203.417--1203.632& 35 & $\mathrm{U}$   &240& 1210.269--1210.391& 151& $\mathrm{R_C}$&60\\
  1203.426--1203.559& 8  & $\mathrm{V}$   &30 & 1211.278--1211.358& 197& $\mathrm{I_C}$&60\\
  1203.428--1203.634& 59 & $\mathrm{R_C}$ &120& 1213.253--1213.339& 155& $\mathrm{I_C}$&60\\
  1203.412--1203.558& 10 & $\mathrm{I_C}$ &30 & 1214.268--1214.348& 195& $\mathrm{I_C}$&60\\
  1205.438--1205.487& 18 & $\mathrm{R_C}$ &120& 1215.250--1215.414& 203& $\mathrm{I_C}$&60\\
  1205.439          & 1  & $\mathrm{I_C}$ &90 & 1216.264--1216.410& 180& $\mathrm{I_C}$&60\\
  1209.527--1209.531& 2  & $\mathrm{B}$   &120& 1217.292--1217.403& 134& $\mathrm{I_C}$&60\\
  1209.528--1209.602& 38 & $\mathrm{V}$   &150& 1221.281--1221.503& 273& $\mathrm{I_C} $&60\\
  1209.525--1209.530& 2  & $\mathrm{R_C}$ &120& \multicolumn{4}{c}{K-380/APOGEE E47} \\
  1209.533--1209.540& 2  & $\mathrm{I_C}$ &90 & 1208.437--1208.606& 149& integral&180\\
  1214.414--1214.533& 7  & $\mathrm{B}$   &45 & 1209.432--1209.582& 133& integral&90\\
  1214.510          & 1  & $\mathrm{U}$   &525&1212.406--1212.597& 169& integral&90\\
  1214.412--1214.532& 7  & $\mathrm{V}$   &30 & 1221.365--1221.537& 81& integral&180\\
  1214.406--1214.532& 10 & $\mathrm{R_C}$ &30 & 1223.301--1223.482& 158& integral&90\\
  1214.429--1214.647& 123& $\mathrm{I_C}$ &150& & & \\
  1215.522--1215.565& 4  & $\mathrm{B}$   &45 & & & \\
  \hline
\end{tabular}
\end{center}
\end{table*}

\section{3. Analysis of the photometry}

\begin{figure*}
  \centering
	\includegraphics[width=0.8\textwidth]{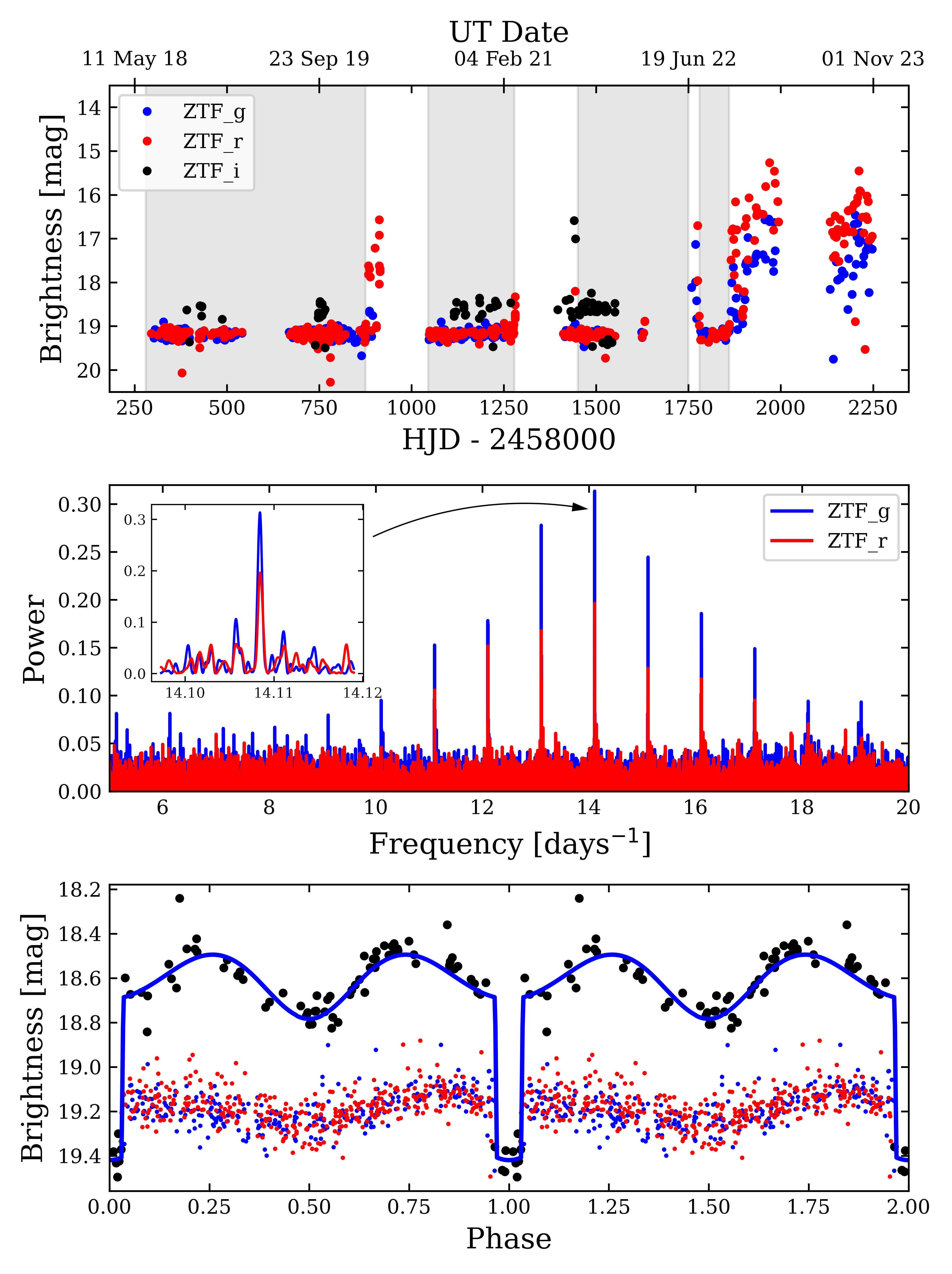}
\caption{Upper panel: the long-term ZTF light curves of Gaia23cer in the $g$, $r$ and $i$ bands (the blue, red, and black dots, respectively). The gray regions indicate the object’s low states. Middle panel: the Lomb–Scargle periodograms constructed from the light curves in the $g$ and $r$ bands (the blue and red lines, respectively) in the low state. Lower panel: the phased light curves in the $g$, $r$ and $i$ bands (the blue, red, and black dots, respectively) in the low state constructed using the ephemeris (\ref{ephem}). The blue line indicates the theoretical light curve calculated for the model of a semidetached binary system (see Section 6). 
}
\label{fig:phot_long}
\end{figure*}

We used ZTF observations \citep{masci18} to analyze the long-term variability of Gaia23cer. Figure \ref{fig:phot_long} shows the light curves of Gaia23cer in the $g$, $r$ and $i$ bands from the ZTF DR20 catalog covering $\approx5.5$~years. They exhibit low ($\langle g \rangle \approx \langle r \rangle \approx 19.2\mathrm{\,mag}$) and high ($\langle r \rangle \sim 16.5\mathrm{\,mag}$) states corresponding to different accretion rates. The same figure presents the Lomb–Scargle periodograms \citep{VanderPlas18} constructed from the low-state data. A power peak at frequency $f = 14.10845 \pm 0.0002$~day$^{-1}$ (period $P = 102.0665 \pm 0.0015$~min) and weaker peaks at $f \pm n $ ($n= 1, 2, \ldots$~day$^{-1}$) attributable to the modulation of the observations with the Earth’s diurnal rotation are distinguished on the periodograms. Figure \ref{fig:phot_long} also presents the phase-foldered light curves of Gaia23cer. Because of the presence of distinct eclipses, it is obvious that this period is orbital. There is a weak out-of-eclipse variability in the $g$ and $r$ bands with an amplitude of $\approx 0.2\mathrm{\,mag}$. In the $i$-band the out-of-eclipse variability has a double-humped structure and a higher amplitude ($\Delta i \approx 0.3\mathrm{\,mag}$). A comparable out-of-eclipse variability amplitude is expected from the donor ellipticity at parameters of the system’s components typical for $P_{orb} \approx 100$~min (for more details, see Section 6).

\begin{figure*}
  \centering
	\includegraphics[width=0.8\textwidth]{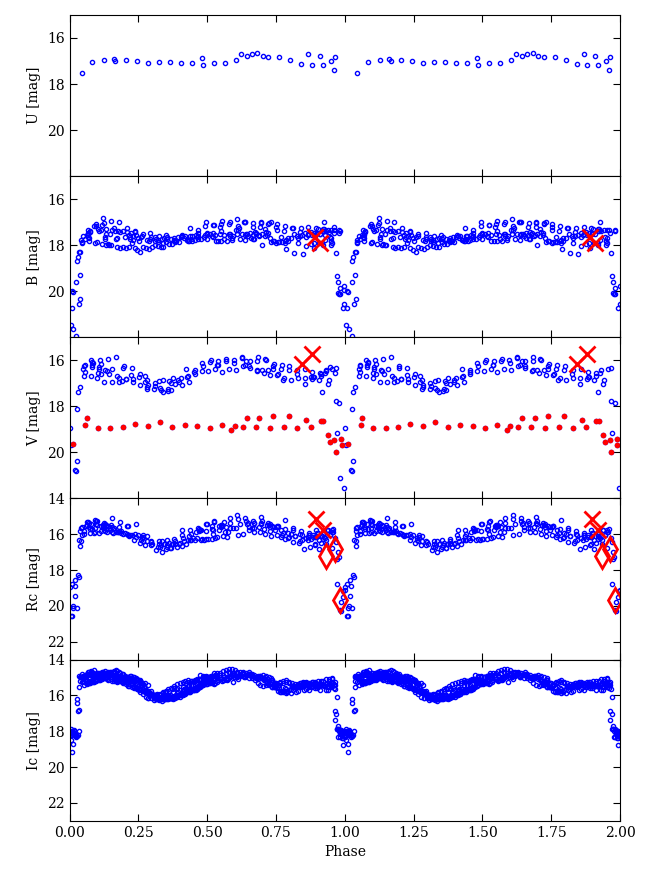}
\caption{The light curves of Gaia23cer constructed from the observations in the fall of 2023 in the $\mathrm{U}$, $\mathrm{B}$, $\mathrm{V}$, $\mathrm{R_C}$ and $\mathrm{I_C}$ bands (blue open circles). The red crosses and red diamonds mark the brightness measurements before the spectroscopic observations on September 12/13, 2023, and October 19/20, 2023, respectively. The red circles mark the observations of Gaia23cer in the low state of January 29/30, 2024. 
}
\label{fig:phot23}
\end{figure*}

The phased light curves of Gaia23cer constructed from our observations in 2023 are shown in Fig. \ref{fig:phot23}. They caught Gaia23cer in its high state with an out-of-eclipse brightness $\mathrm{V} = 16-17\mathrm{\,mag}$. Outside eclipses the light curves in the $\mathrm{V}$, $\mathrm{R_C}$, and $\mathrm{I_C}$ bands have a double-humped shape with a separation between the humps of $\approx {^1/_2} P_{orb}$ and brightness amplitudes  $\Delta \mathrm{V} \approx 1\mathrm{\,mag}$, $\Delta \mathrm{R_C} \approx 1.2\mathrm{\,mag}$ and $\Delta \mathrm{I_C} \approx 1.3\mathrm{\,mag}$. Remarkably, there is no plateau phase with a reduced brightness and an extent of $\approx {^1/_2} P_{orb}$ in the light curves. For an eclipsing system (i.e., a system with a high inclination of the white dwarf rotation axis $i\sim 90^{\circ}$) this phase would have been expected in the case of a single accretion spot, when it is located behind the observed white dwarf disk (see, e.g., \cite{Kolbin22}). The absence of the plateau phase may point to two-pole accretion in the high state, since in the case of two or more spots at least one of them will be visible to the observer at an arbitrary orbital phase. Figure \ref{fig:phot23} also shows the light curve of Gaia23cer in the low state of January 29/30, 2024, in the $\mathrm{V}$ band. In contrast to the high state, the out-of-eclipse variability is less distinct and does not exceed $\Delta \mathrm{V} = 0.3\mathrm{\,mag}$. The mid-eclipse ephemeris was obtained through a joint analysis of the ZTF, ZTSh, AZT-11, and Zeiss-600 data:
\begin{equation}
    \mathrm{HJD_{min}} = 2460201.43461(17) + 0.070879652(9) \times E.
\label{ephem}
\end{equation}

\begin{figure}
  \centering
	\includegraphics[width=0.9\linewidth]{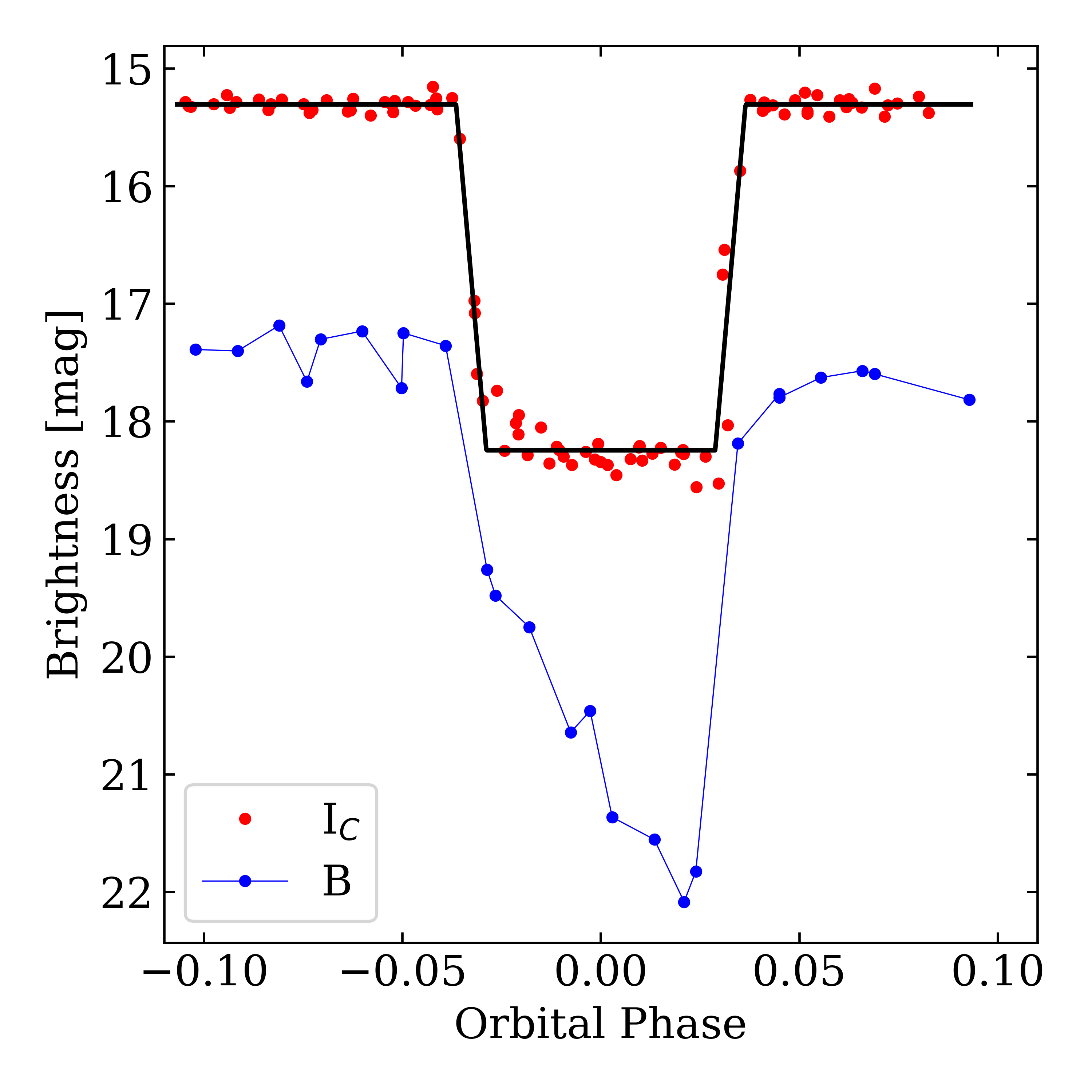}
\caption{The light curves of Gaia23cer near the eclipse in the $\mathrm{B}$ and $\mathrm{I_C}$ bands. The black line is the fit to the eclipse profile by the trapezoid. 
}
\label{fig:ecl}
\end{figure}

Figure \ref{fig:ecl} shows the Gaia23cer eclipse profiles obtained with the ZTSh telescope in the $\mathrm{B}$ and $\mathrm{I_C}$ bands. In the $\mathrm{B}$ band there is an asymmetry of the eclipse profile with a gradual ingress and a rapid egress. Such a behavior of the brightness in eclipse is typical for polars in a high state and is interpreted by the occultation of the bright accretion stream by the donor (see, e.g., \cite{Rodriguez23, Borisov16}). In the $\mathrm{I_C}$ band the ratio of the accretion stream luminosity to the total luminosity of the accretion spot and the white dwarf is smaller, which shows up as a weaker manifestation of the stream in the eclipse profile (a similar attenuation of the flux from the stream with wavelength was detected, for example, in HU Aqr by \cite{Harrop99}). To estimate the duration of the eclipse, its profile in the band was fitted by a trapezoid. The width of the trapezoid at half depth $\Delta t_{ecl} = 401.30 \pm 0.81$~s was taken as the white dwarf eclipse duration. It is difficult to measure the duration of the eclipse ingress $\Delta t_{ing}$ (and the duration of the eclipse egress equal to it) due to the long exposure time that smears the eclipse profile. However, from the fit of the light curve by the trapezoid we can impose the constraint $\Delta t_{ing} < 47 \pm 2 $~s. The errors in the eclipse parameters were found by the Monte Carlo simulations and correspond to a scatter of $1\sigma$.

\section{4. Spectral analysis}

\begin{figure*}
  \centering
	\includegraphics[width=\linewidth]{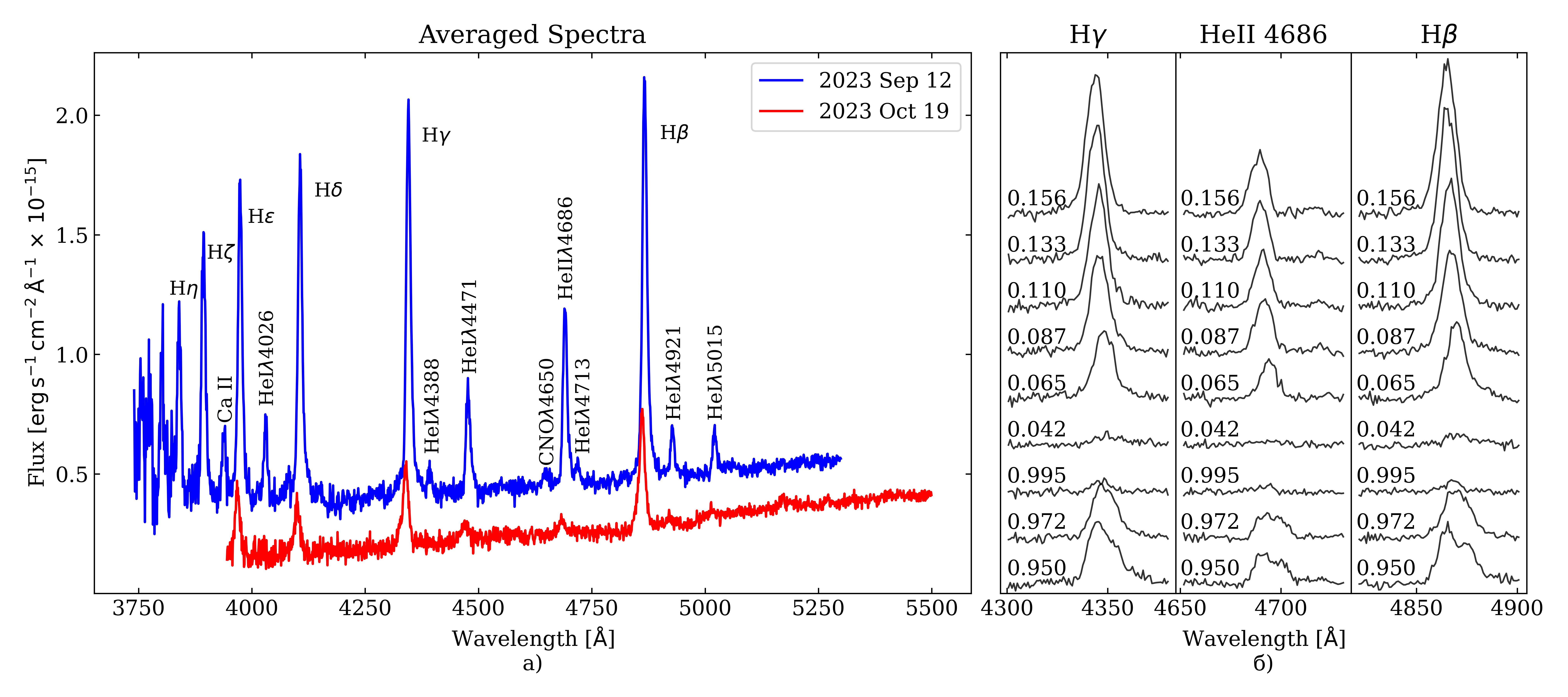}
\caption{
(a) The averaged spectra of Gaia23cer from the September 12/13, 2023 and October 19/20, 2023 data. (b) The evolution of the H$\beta$, H$\gamma$ и HeII~$\lambda4686$ line profiles near the eclipse phase.}
\label{fig:spectra}
\end{figure*}

The averaged spectra of Gaia23cer from the September 12/13 and October 19/20 data (for the VPHG1200G grism) are shown in Fig. \ref{fig:spectra}. The spectra contain a set of emission lines typical for polars: Balmer hydrogen lines, neutral helium lines, and the  HeII~$\lambda4686$ line. The great difference in the residual intensity of the  HeII~$\lambda4686$ line between the two sets of spectroscopic observations engaged our attention: in the October 19/20 observations this line almost disappears, whereas on September 12/13 its intensity is approximately half of the H$\beta$ intensity. This phenomenon can be associated with the transition of the polar to its low state on October 19/20. Indeed, the SCORPIO--1 photometric observations performed immediately before the spectroscopic observations show a decrease in the brightness of Gaia23cer by $\Delta \mathrm{R_C} = 1-1.5\mathrm{\,mag}$ on the night of October 19/20 compared to the night of September 12/13 (see Fig. \ref{fig:phot23}).

Figure \ref{fig:spectra}b shows the set of H$\beta$, H$\gamma$ and HeII~$\lambda$4686 line profiles obtained near the eclipse phase on September 12/13. The eclipse of the source of emission features lags behind the eclipse of the white dwarf by  $\Delta \varphi =0.04\pm0.02$. This can be seen from a comparison of the spectra taken at phases $\varphi=0.972$ and $\varphi=0.042$. Although the first spectrum was taken closer to the center of the white dwarf eclipse ($\varphi=0$) than the second one, it exhibits much stronger lines (only faint traces of the lines are seen at $\varphi=0.042$). The noted phenomenon is natural when the lines are formed in the accretion stream. The stream outflowing from the Lagrange point L$_1$ is known to be deflected from the line connecting the centers of the stars of the binary system by the Coriolis force in the direction of orbital motion of the donor. Accordingly, the eclipse of the stream by the donor occurs later than the eclipse of the white dwarf. A double-peaked structure of the H$\beta$ and HeII~$\lambda$4686 line profiles at $\varphi = 0.95$ is also seen in Fig. \ref{fig:spectra}b. This phenomenon is probably caused by the absorption feature near the line center that resulted from the occultation of the accretion spot by an optically thick accretion stream. Such a phenomenon is typical in eclipsing polars (see, e.g., \cite{Kochkina23, Rodriguez23, Borisov16}).

\begin{figure}
  \centering
	\includegraphics[width=\linewidth]{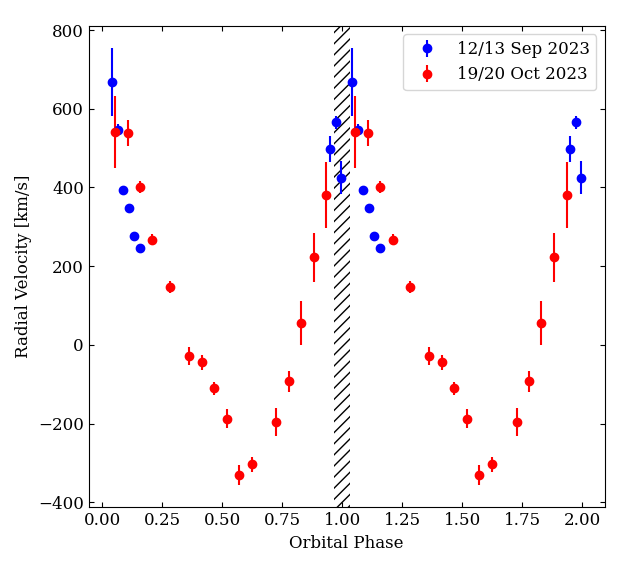}
\caption{
The radial velocity curve of Gaia23cer obtained from the September 12/13, 2023 and October 19/20, 2023 observations. The hatched region indicates the time interval corresponding to the eclipse. }
\label{fig:rvs}
\end{figure}

The radial velocities of Gaia23cer were determined by Gaussian fitting of H$\beta$ line which has the best signal-to-noise ratio in the two series of spectra. The radial velocity curve constructed from the data of the two nights was folded with the orbital period and is shown in Fig. \ref{fig:rvs}. The radial velocities are seen to be modulated with the orbital motion and to have a high semi-amplitude $K\approx 450$~km/s typical of AM Her-type variables. The radial velocity reaches its maximum near an orbital phase $\varphi \approx 0$, which is also expectable for AM Her-type systems (see, e.g., \cite{Borisov16}).

\begin{figure*}
  \centering
	\includegraphics[width=\textwidth]{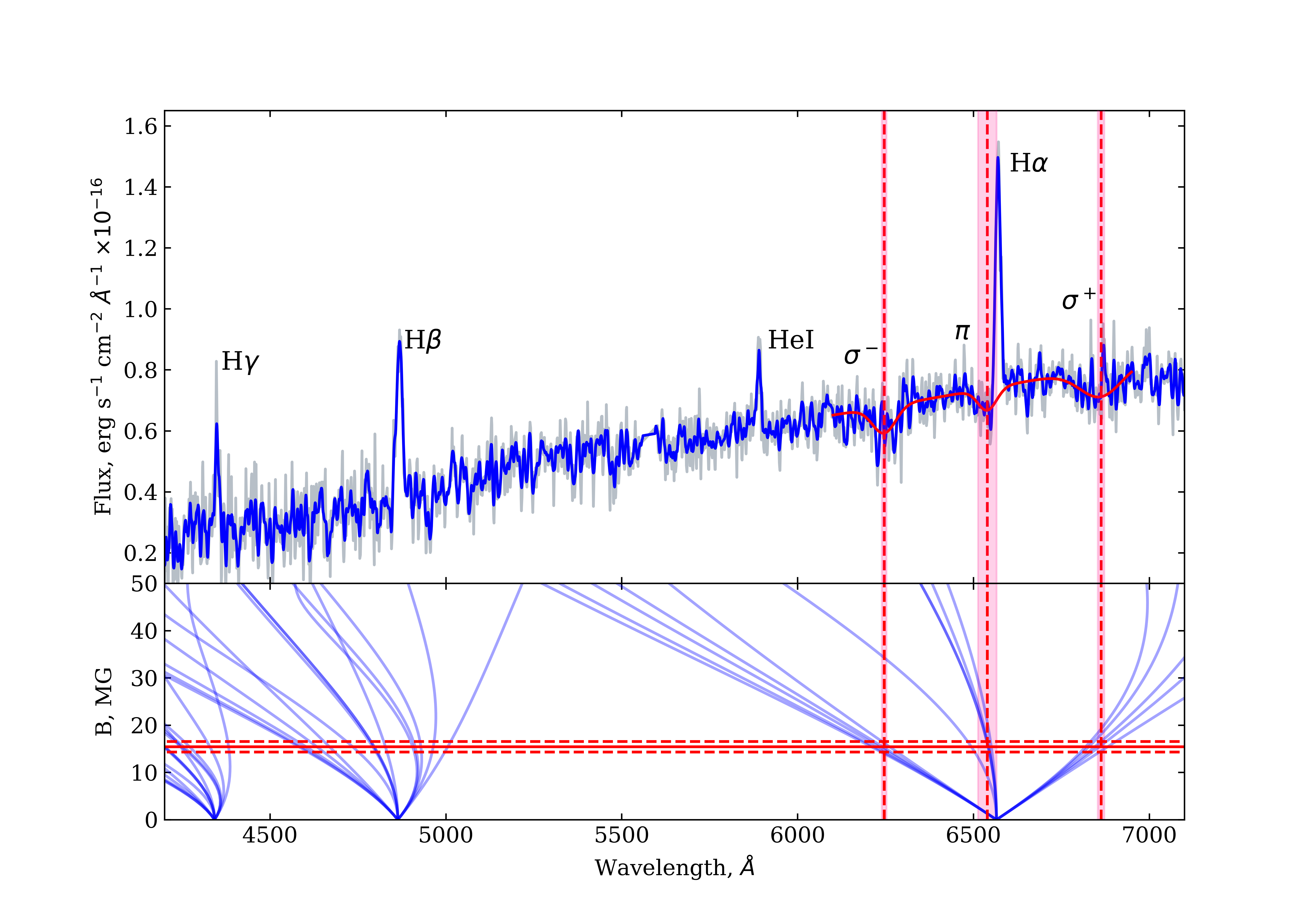}
\caption{
Top panel: the observed spectrum of Gaia23cer (gray line), the smoothed spectrum obtained by the Savitsky--Golay filter (blue line), and the fit of the Н$\alpha$ Zeeman splitting components by the sum of a low-degree polynomial and three Gaussians (red line). Bottom panel: the wavelengths of the Zeeman splitting components of the H$\alpha$, H$\beta$ and H$\gamma$ lines for the range of magnetic field strengths $0-50$~MG. The vertical lines indicate the positions of components of the Zeeman H$\alpha$ triplet; the horizontal lines indicate the estimate of the magnetic field and its uncertainty. 
}
\label{fig:zeeman}
\end{figure*}

Figure \ref{fig:zeeman} shows the averaged spectrum of Gaia23cer obtained with the VPHG550G grism. It covers the $4200-7100$~\AA\, range and is fairly noisy due to slight cloudiness at the time of observations. The spectrum has a red cyclotron continuum without distinguishable cyclotron harmonics. The spectrum contains weak absorption Zeeman $\sigma^-$, $\pi$ and $\sigma^+$ components of the Н$\alpha$ line. From the Н$\alpha$ splitting diagram shown in Fig. \ref{fig:zeeman}, the magnetic field strength is $B = 15.2\pm1.1$~MG. Note that the telluric Fraunhofer line B was removed in the presented spectrum. The line was removed by dividing the polar’s spectrum by the normalized spectrum of the standard star. The wavelengths of the Zeeman splitting components of the Balmer lines were calculated using the code of \cite{Schimeczek14}.

\begin{comment}
Интересны  фазовые кривые в высоком и низком состоянии. Амплитуда кривой в инфракрасной области уменьшилась от $\sim$ 3 маг (высокое состояние) (\ref{fig:23cerphase1} до  $\sim$ 1 маг (низкое состояние). И профиль кривых в более коротковолновых областях изменился.
\end{comment}

\section{5. The white dwarf}

\begin{figure*}
  \centering
	\includegraphics[width=0.8\textwidth]{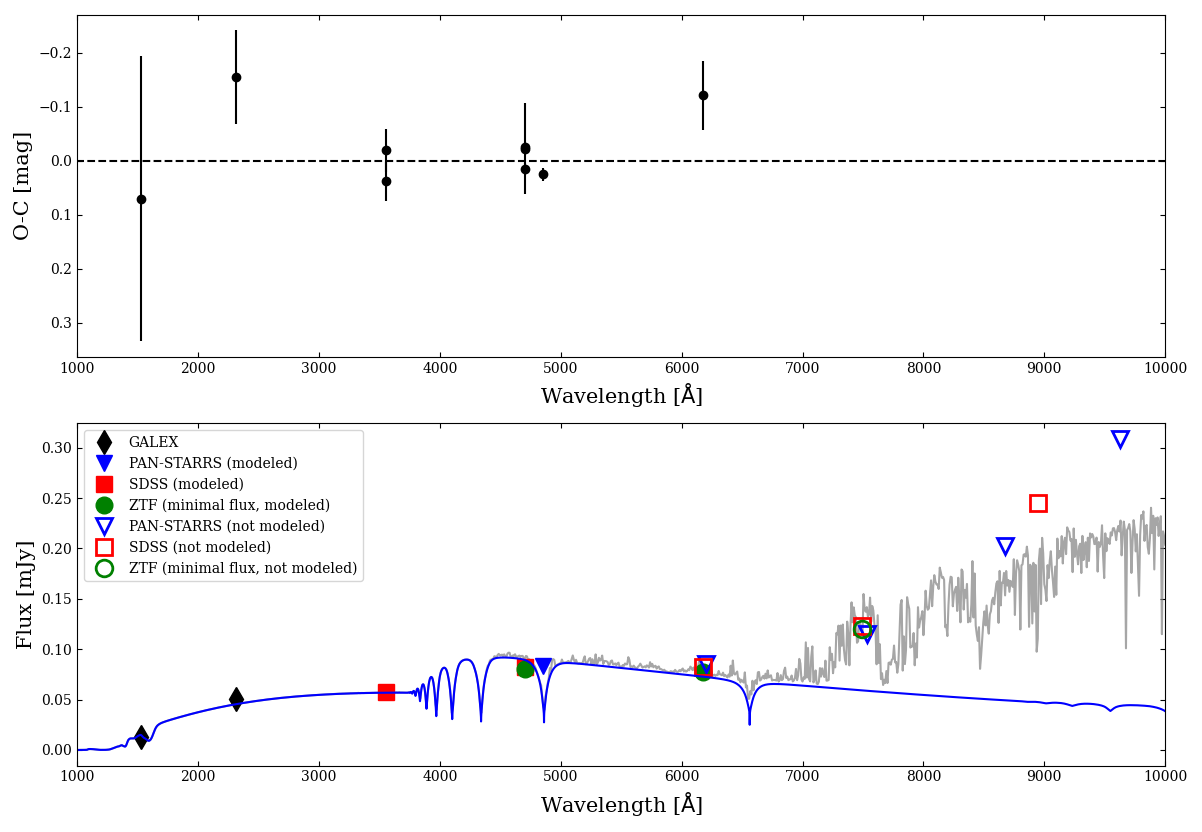}
\caption{
Bottom panel: the spectral energy distribution of Gaia23cer, the synthetic spectrum of the white dwarf (blue line) and the sum of the spectra of the white dwarf and the donor (gray line). Top panel: the $O-C$  diagram for the modeled fluxes in the ultraviolet and optical ranges. 
}
\label{fig:sed}
\end{figure*}

We reconstructed the spectral energy distribution of Gaia23cer based on archival GALEX observations \citep{Boselli11} in the FUV ($\lambda_{\mathrm{eff}} \approx 1549$~\AA) and NUV ($\lambda_{\mathrm{eff}} \approx 2303$~\AA) bands, SDSS data (the $u$, $g$, $r$ and $i$ bands \cite{Ahumada20}), and Pan-STARRS data \citep{Chambers16}. We added the fluxes from the ZTF survey corresponding to the polar’s low state to them. According to the STELISM three-dimensional interstellar extinction maps\footnote{https://stilism.obspm.fr, for more details see \cite{extmap1, extmap2, extmap3}.}, the color excess for Gaia23cer is $E(B-V) = 0.024\pm0.017\mathrm{\,mag}$ with a corresponding extinction $A_V = 3.1 E(B-V) = 0.074 \pm 0.053\mathrm{\,mag}$. The observed fluxes from Gaia23cer were corrected for interstellar extinction using the extinction curve from \cite{Fitzpatrick99}. The derived spectral energy distribution of Gaia23cer is shown in Fig. \ref{fig:sed}. It can be seen from a comparison of the ZTF fluxes with the SDSS and PAN-STARRS fluxes that the latter were measured in the low state. Owing to the relatively weak magnetic field of Gaia23cer, the main emission source in the ultraviolet must be the white dwarf. It can also be assumed that the white dwarf dominates in the optical fluxes in the low state.

\begin{figure}
  \centering
	\includegraphics[width=\linewidth]{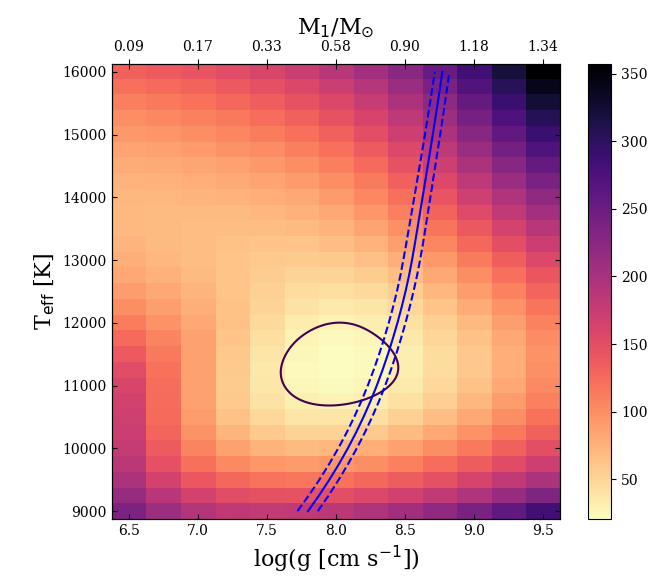}
\caption{
The $\chi^2$ map in the $T_{eff} - \log g$ plane obtained when modeling the spectral energy distribution of Gaia23cer by the white dwarf model. The black closed line corresponds to the 95\% confidence level. The blue line represents the $T_{eff} - \log g$ relation deduced from the correspondence of the angular diameter of the white dwarf to its atmospheric parameters (for more details, see the text). 
}
\label{fig:chi2map}
\end{figure}

We attempted to reproduce the spectral energy distribution of Gaia23cer based on the spectra of hydrogen-rich (DA) white dwarf model atmospheres computed by \cite{Koester10}. The theoretical spectra were interpolated in effective temperature $T_{eff}$ and  surface gravity $\log g$ when calculating the fluxes for given stellar parameters. The theoretical fluxes in the photometric bands were calculated by convolving the white dwarf spectrum with the corresponding transmission functions\footnote{Функции пропускания использованных фильтров доступны на сайте http://svo2.cab.inta-csic.es/theory/fps/.}. The parameters of the white dwarf were determined by minimizing the weighted sum of the residuals between the observed fluxes in photometric bands, $F$, and their theoretical estimates, i.e.,
\begin{equation}
    \chi^2 = \sum_i \Big(\frac{F_i - \theta^2 \hat{F}_i/4 }{\sigma_i}\Big)^2,
\end{equation}
where $\theta$ is the angular diameter of the white dwarf, $\hat{F}$ is the theoretical flux near the stellar surface, and $\sigma$ is the flux measurement error. For the free parameter $\theta$ the minimum of $\chi^2$ is reached at $T_{eff}=11350\pm650 K$ and $\log g = 7.60-8.47$ (the scatter of parameters corresponds to the 95\% confidence level). The $O-C$ diagram for the best solution is presented in Fig. \ref{fig:sed}. The same figure compares the estimates of the observed fluxes at the central wavelengths of the photometric filters with the theoretical spectrum of the white dwarf. A map of the $\chi^2$ distribution with the 95\% confidence region is presented in Fig. \ref{fig:chi2map}.

The atmospheric parameters of the white dwarf presented above were found from the approximation of the shape of the spectral energy distribution . In this case, the angular diameter was considered as a free parameter. An improved value of $\log g$ can be found from the relation of the angular diameter $\theta$ to the white dwarf radius $R_1$, which is related to the mass as
\begin{equation}
    R_1 = 0.78\times10^9 \mathrm{cm}\Big[\Big(\frac{M_{ch}}{M_1}\Big)^{\frac{2}{3}}-\Big(\frac{M_1}{M_{ch}}\Big)^{\frac{2}{3}}\Big]^{1/2},
\end{equation}
where $M_{ch} = 1.44 M_{\odot}$ \citep{Nauenberg72}. The requirement for the calculated angular diameter of the white dwarf to be consistent with its radius can be formulated in the equation
\begin{equation}
    2\frac{R_1(\log g)}{D} = \argmin_{\theta} \chi^2(\theta, \log g, T_{eff}).
\end{equation}
Thus, knowing the distance to the object $D$, we can solve the last equation for $\log g$ in a set of temperatures. This gives us yet another solution in the $T_{eff} - \log g$ plane shown in Fig. \ref{fig:chi2map}. According to the Gaia DR3 catalog, the parallax of Gaia23cer is $p'' = (4.3261 \pm 0.2)\times10^{-3}$ (the distance is $D=231 \pm 11$~pc). Within the confidence interval of the temperatures the angular diameters correspond to $\log g = 8.26-8.38$ or a white dwarf mass $M_1 = 0.79\pm 0.03 M_{\odot}$. A radius $R_1 = 0.0101 \pm 0.0003R_{\odot}$ corresponds to this mass. Note that the duration of the eclipse ingress for a white dwarf with such a radius is $\Delta t_{ing} \approx 30$~s (for an orbital inclination $i=85^{\circ}$, see the next section). This value is consistent with the constraint on the duration of the eclipse ingress $\Delta t_{ing} < 47$ found by analyzing the photometry for Gaia23cer (see Section 3).

\section{6. The donor}

\begin{figure}
  \centering
	\includegraphics[width=\linewidth]{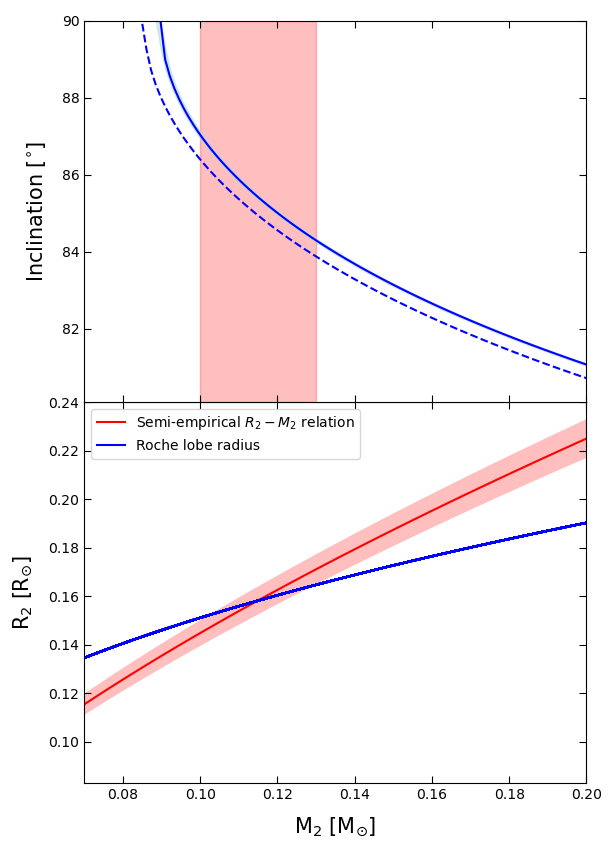}
\caption{
Upper panel: the solution in the $i-M_2$ plane that provides the Gaia23cer eclipse duration (blue line). The dashed line indicates the solution corresponding to the eclipse of the point on the white dwarf surface (accretion spot) closest to the donor. The red region indicates the range of possible donor masses obtained from the requirement for the Roche lobe radius to be equal to the evolutionary radius (see the lower panel). Lower panel: the semi-empirical mass–radius relation for the donors in cataclysmic variables (red line) and the effective donor Roche lobe radius $R_L$--mass relation (blue line). The radius $R_L$ depends weakly on the accretor mass $M_1$ at a fixed orbital period $P_{orb}$, while the presented $R_L(M_2)$ curve is a family of indistinguishable solutions for $M_1 \in 0.4-1.44 M_{\odot}$. 
}
\label{fig:m2solution}
\end{figure}

The donor mass can be constrained from the eclipse duration. Upper panel of the Fig. \ref{fig:m2solution} presents the solution in the $i-M_{2}$ plane ($i$ is the orbital inclination) that provides the observed eclipse duration $\Delta t_{ecl} = 401.30 \pm 0.81$ s. When constructing it, we assumed the shape of the donor to be described by the surface of its Roche lobe. It follows from this figure that the donor mass must be $\geq 0.09M_{\odot}$. The presented solution was obtained from the assumption of the white dwarf eclipse, i.e., it corresponds to the occultation of its center. In reality, because of the high luminosity of the accretion spot, the eclipse profile in the high state can reflect the occultation of the accretion spot. It is difficult to obtain the solution for the eclipse of the accretion spot due to the absence of information about its coordinates on the stellar surface. However, to estimate the uncertainties caused by the contribution of the accretion spot, in Fig. \ref{fig:m2solution} we displayed the solution corresponding to the eclipse of a point on the white dwarf surface facing toward the donor. We see that the donor mass and the orbital inclination may be overestimated by $\Delta M_2 = 0.004 M_{\odot}$ and $\Delta i = 0.5^{\circ}$, respectively.

Another constraints on the system’s parameters can be imposed from the semi-empirical mass–radius relation for the donor in cataclysmic variables given by \cite{McAllister19}  (see also \cite{Knigge11}). In the range of masses $M_2 \in 0.063-0.2 M_{\odot}$ this relation is fitted by a power law:
\begin{equation}
    \frac{R_2}{R_{\odot}} = 0.225 \pm 0.008 \Big(\frac{M_2}{0.2M_{\odot}}\Big)^{0.636\pm0.012},
\end{equation}
which is shown \ref{fig:m2solution} (lower panel). On the other hand, the effective Roche lobe radius of the donor $R_L$ (i.e., the radius of the sphere whose volume is equal to the volume of the Roche lobe) can be estimated as
\begin{equation}
R_L=A\frac{0.5126 q^{0.7388}}{0.6710q^{0.7349}+\mathrm{ln}(1+q^{0.3983})},
\label{eq_rl}
\end{equation}
where $A=(M_2(1+1/q)P_{orb}^2)^{1/3}$ is the semimajor axis of the system and $q=M_2/M_1$ is the component mass ratio. Formula (\ref{eq_rl}) is valid for fully convective stars described by polytropic models with an index $n=3/2$ \citep{Sirotkin09, Knigge11}. Since our system is below the period gap ($P_{orb}<2$~ч), it must contain a fully convective donor \citep{Knigge11}. It follows from Fig. \ref{fig:m2solution} that the effective Roche lobe radius is consistent with the semi-empirical $R_2-M_2$ relation in the range of masses $M_2 \in 0.10-0.13 M_{\odot}$ and the range of radii $R_2 \in 0.152 - 0.164R_{\odot}$. Returning to the solution in the $i-M_2$ plane, we see that an orbital inclination $84.3^{\circ} \le i\le 87.0^{\circ}$ corresponds to this range of masses.

In the $i$ band in the low state the contribution from the donor to the total flux from the system is significant. This is suggested by the double-humped light curve shown in Fig. \ref{fig:phot_long} (lower panel). Such a behavior of the brightness is expected for the ellipticity of the secondary component. We modeled this light curve by the model of a semidetached binary system using the PHOEBE\footnote{The PHOEBE code is accessible at http://www.phoebe-project.org.} \citep{Prsa16}. Our modeling was performed for the white dwarf parameters found in the previous section ($M_1 = 0.79M_{\odot}$, $R_1 = 0.01R_{\odot}$, $T_{eff} = 11350K$). The donor mass was taken to be $0.12M_{\odot}$, and the orbital inclination was fixed at $i=85^{\circ}$. We calculated the intensity of the radiation from both components in the blackbody approximation and took into account the limb darkening for the donor within the linear law (we chose the blackbody approximation, because we went beyond the grid of model atmosphere parameters used in PHOEBE near the Lagrange point L$_1$). The limb darkening parameters were found by interpolating the tables from \cite{Claret12} computed for the PHOENIX model atmospheres. The best fit of the observations was achieved for a donor temperature $T\approx 2850~K$. The observed and model light curves are compared in Fig. \ref{fig:phot_long}. 

The brightness of Gaia23cer in eclipse in the low state is $i = 19.41 \pm 0.02\mathrm{\,mag}$ and can be used to estimate the donor temperature. We computed the model fluxes near the stellar surface in the $i$ band by convolving the synthetic spectra of cool stars BT-Settl (CIFIST) \citep{Allard03, Allard11} with the transmission function of the photometric filter. The chemical composition of the donor atmosphere was assumed to be solar. Agreement between the observed and model fluxes at a distance to Gaia23cer $D=231\pm11$~pc and a donor radius $R \in 0.152-0.164 R_{\odot}$ is achieved for $T_{eff} = 2900 \pm 40 K$. This temperature estimate is close to the temperature found by modeling the light curve of Gaia23cer in the low state. Figure \ref{fig:sed} compares the spectral energy distribution of Gaia23cer in the low state with the combined spectrum of the white dwarf and the donor. The donor is seen to make a minor contribution to the $g$ and $r$ bands (no more than 8\%). In the $z$ and $y$ bands the flux from the stellar components is not enough to describe the observed flux. Possibly, for wavelengths $\lambda \gtrsim 9000$~\AA\, the contribution of the cyclotron radiation is significant even in the low state.

\section{CONCLUSIONS}

In this paper we performed photometric and spectroscopic study of the new eclipsing polar Gaia23cer. Below we summarize our main conclusions.

Low and high states differing in brightness by $\Delta r \approx 2.5\mathrm{\,mag}$ and attributable to the change in accretion rate are distinguished in the long-term light curves of Gaia23cer. The orbital period of the polar is $P_{orb} = 102.0665 \pm 0.0015$ min. There are deep eclipses with a duration of $401.30 \pm 0.81$ s. In the hight state the eclipse profile has an asymmetry caused by the eclipse of the accretion stream. In the same state the optical light curves have a double-humped out-of-eclipse variability probably attributable to two-pole accretion. In the low state the contribution of the donor in the $i$ band is significant, which manifests itself as in ellipticity effect.

The spectra of Gaia23cer exhibit a red cyclotron continuum without any distinguishable cyclotron harmonics that is often encountered in polars with a low magnetic field $B\lesssim20$~МГс \citep{Schwope95, Schwope97, Kolbin23}. The magnetic field strength $B = 15.2 \pm 1.1$ MG estimated from the Zeeman H$\alpha$ absorption triplet points to a weak magnetic field of the white dwarf. It should be noted that, in contrast to the H$\alpha$ line, the Zeeman H$\beta$ and H$\gamma$ components are not observed on more than a factor of 2 weaker cyclotron continuum. Thus, the Zeeman H$\alpha$ splitting is probably formed in the cold halo surrounding the accretion spot (\citep{Achilleos92}; see also \cite{Schwope95, Kolbin23}). The behavior of the emission lines is typical for AM Her systems. The radial velocities are modulated with the orbital period and have a high semi-amplitude $K\approx 450$~km/s. There is a noticeable difference in the eclipse times of the white dwarf and the emission line source by $\Delta \varphi \approx 0.04$ expected when the lines are formed in the accretion stream.

The atmospheric parameters of the white dwarf, $T_{eff}=11350\pm650 K$ and $\log g = 8.26-8.38$, were estimated by modeling the spectral energy distribution in the ultraviolet and optical ranges. A temperature $T\sim 10000 K$ is typical for white dwarfs in cataclysmic variables below the period gap $P_{orb} \lesssim 2$~h \citep{Townsley09, Pala22}. A white dwarf mass $M_1 = 0.79 \pm 0.03 M_{\odot}$ and radius $R_1 \approx 0.01~R_{\odot}$ correspond to the above range of surface gravities. The derived mass is close to the mean white dwarf mass in cataclysmic variables $\langle M_1 \rangle = 0.83 \pm 0.23 M_{\odot}$ \citep{Zorot11}.

The eclipse duration in Gaia23cer gives a constraint on the donor mass $M_2 \ge 0.09 M_{\odot}$. Thus, the donor mass in Gaia23cer exceeds the limiting mass $M_{bounce} = 0.063^{+0.005}_{-0.002} M_{\odot}$ \citep{McAllister19} starting from which the evolution of the system changes from a decreasing period to an increasing one. It follows from the requirement for the evolutionary donor radius to be equal to the effective Roche lobe radius that the donor mass must lie in the range $M_2 \in 0.10-0.13 M_{\odot}$. The eclipse duration for this range of masses corresponds to the range of orbital inclinations $i \in 84.3- 87.0^{\circ}$. The donor temperature was estimated by modeling the $i$-band light curve in the low state to be $T \approx 2850K$. The donor temperature was also determined from the brightness in eclipse and the known distance to Gaia23cer. The estimate of $T=2900 \pm 40 K$ obtained in this way is consistent with the result light curve modeling. Note that the temperature $T=2900 K$ is expectable for the evolutionary status of Gaia23cer, in which the system, loosing its angular momentum, has traversed the period gap ($2$~h $\lesssim P_{orb}\lesssim 3$~h), but has not yet reached its minimal period $P_{\mathrm{min}} \approx 76 - 82$~min \citep{Knigge11, McAllister19}.

In conclusion, note that the possibility of reliably determining the parameters of the components in Gaia23cer and, particularly, such evolutionary indicators as the white dwarf temperature and the donor radius makes it an interesting object for understanding the evolution of binary systems with magnetized components. In addition, owing to its fairly high brightness ($r=16.5-18.5\mathrm{\,mag}$ outside eclipses) and the existence of eclipses, Gaia23cer is an interesting object for studying the magnetically controlled accretion by the Doppler \citep{Marsh16}, Stokes \citep{Kolbin20}, and eclipse \citep{Harrop99, Harrop01} mapping techniques.

{\bf Funding} 
The observations of A.A. Sosnovskij were supported by grant no. 23-72-01080 of the Russian Science Foundation (https://rscf.ru/project/23-72-01080). The work of A.I. Kolbin, V.Yu Kochkina and M.V. Suslikov (analysis of spectra and photometry, determination of binary parameters and physical parameters of components) was supported by grant no. 22-72-10064 of the Russian Science Foundation (https://rscf.ru/project/22-72-10064/). The observations of S.Yu. Shugarov at the Zeiss-600 telescope (Tatranska Lomnica, Slovakia) were supported by APVV grant no. 20-0140 of the Slovak Agency for Research and Development and VEGA grant no. 2/0030/21 of the Slovak Academy of Sciences. The observations with the SAO RAS telescopes are supported by the Ministry of Science and Higher Education of the Russian Federation. The instrumentation is updated within the ‘‘Science and Universities’’ National Project.

\end{document}